\def \malpunkt· 
\def \zl #1 #2;{\ensuremath{#1 \thinspace \textrm{#2}}}
\def \zle #1 #2 #3;{\def \test{#3}\ensuremath{\if #11\else #1 \malpunkt \fi 10^{#2} \ifx \test\empty \else \thinspace \textrm{#3} \fi}}
\def \zlr #1 #2 #3;{\ensuremath{(#1 {\thinspace\pm\thinspace} #2) \thinspace \textrm{#3}}}
\def \zler #1 #2 #3 #4;{\def \test{#4}\ensuremath{(#1 {\thinspace \pm \thinspace} #3) \malpunkt 10^{#2} \ifx \test\empty \else \thinspace \textrm{#4} \fi}}
\documentclass[preprint,showpacs,preprintnumbers,amsmath,amssymb]{revtex4-1}

\usepackage{graphicx}
\usepackage{dcolumn}
\usepackage{bm}
\usepackage[latin1]{inputenc}

\begin{document}

\preprint{}
\title{Origin of training effect of exchange bias in Co/CoO due to irreversible thermoremanent magnetization of the magnetically diluted antiferromagnet}

\author{S. R. Ali}
\email{rizwan@physik.rwth-aachen.de }
\author{M. R. Ghadimi}
\author{M. Fecioru-Morariu}
\author{B. Beschoten}
\author{G. Güntherodt}
\affiliation{II. Institute of Physics, RWTH Aachen University, 52056
Aachen, Germany}
\date{\today}

\begin{abstract}

The irreversible thermoremanent magnetization ($m_{\rm{TRM}}^{irr}$)
of a sole, magnetically diluted epitaxial antiferromagnetic
Co$_{1-y}$O(100) layer is determined by the mean of its
thermoremanent magnetizations ($m_{\rm{TRM}}$) at positive and
negative remanence. During hysteresis-loop field cycling,
$m_{\rm{TRM}}^{irr}$ exhibits successive reductions, consistent with
the training effect (TE) of the exchange bias measured for the
corresponding Co$_{1-y}$O(100)/Co($11\bar{2}0$) bilayer. The TE of
exchange bias is shown to have its microscopic origin in the TE of
$m_{\rm{TRM}}^{irr}$ of the magnetically diluted AFM.

\end{abstract}


\maketitle

The phenomenon of exchange bias (EB) originates from the interfacial
exchange coupling between an antiferromagnet (AFM) and a ferromagnet
(FM).\cite{1,2,3} This interaction results for the magnetic
hysteresis loop of the FM layer in a field offset from the origin by
the EB field, $B_{\rm{EB}}$. EB has been in the focus of intense
research activities because of its potential applications in
spintronics devices where it stabilizes a reference FM magnetization
in magnetic read heads, sensors and nonvolatile memory devices
\cite{4,5}. It has been shown experimentally that field cooling of
an AFM stabilizes pinned uncompensated moments near the AFM/FM
interface, which are responsible for the EB
effect.\cite{6,peter,7,8,9,10} A domain state develops upon field
cooling of the AFM, which carries an irreversible surplus
thermoremanent magnetization, $m_{\rm{TRM}}^{irr}$. The crucial role
of $m_{\rm{TRM}}^{irr}$ at the AFM/FM interface for the EB effect
has been demonstrated both experimentally \cite{7,8,9,10} and by
Monte Carlo simulations.\cite{11} At the surface and in the bulk of
the AFM there may be structural and substitutional defects
\cite{12}, giving rise to domain wall pinning and thus leading to
metastable domain structures whose evolution with field cycling is
responsible for the training effect (TE). The latter is a crucial
feature associated with the fundamentals and applications of EB due
to the reduction in $B_{\rm{EB}}$ during successive field cycles in
hysteresis loops.\cite{1,3} The TE plays an essential role in the
reliable performance of devices based on EB. The microscopic origin
of the TE remains under intensive debate (see, e.g., Refs.
\cite{1,2,3, 11, 13,14,15,16,17}) and raises the question about the
involvement of, e.g., $m_{\rm{TRM}}^{irr}$ at the AFM/FM interface.
However, the smallness of $m_{\rm{TRM}}^{irr}$ \cite{18,19} remains
a serious difficulty in answering this question.\cite{20} A simple
approach might be to consider a sole AFM layer with a dilution
enhanced $m_{\rm{TRM} }$, i.e. $m_{\rm{TRM}}^{irr}$, such that its
role for the TE could unambiguously be investigated by magnetometry.

Here, we utilize nonmagnetic dilution throughout the bulk of an
epitaxially grown Co$_{1-y}$O(100) layer ($y\rightarrow0$) to
significantly enhance its $m_{\rm{TRM} }$. This in turn also yields
an enhanced $B_{\rm{EB}}$ for the corresponding
Co$_{1-y}$O(100)/Co($11\bar{2}0$) bilayer. The $m_{\rm{TRM}}^{irr}$
of a sole AFM layer is then determined by the difference of its
enhanced $m_{\rm{TRM} }$ at positive and negative remanence. The
measured $m_{\rm{TRM}}^{irr}$ exhibits systematic reductions during
successive field cycling. Detailed analysis of the data using
Binek's model \cite{13} shows that the TE of $B_{\rm{EB}}$ of the
AFM/FM bilayer has its origin in the TE of $m_{\rm{TRM}}^{irr}$ of
the sole AFM.

Diluted ($y\neq0$) and undiluted ($y\rightarrow0$) sole epitaxial
AFM samples with the layer sequence: MgO(100)/Co$_{1-y}$O(100)/Au(5
nm) and epitaxial AFM/FM bilayers with the layer sequence:
MgO(100)/Co$_{1-y}$O(100)/Co($11\bar{2}0$)/Au(5 nm) were deposited
by molecular beam epitaxy (MBE) on MgO(100) substrates. The samples
were capped by a 5 nm thick Au layer and the thicknesses of CoO and
Co are 30 nm and 8 nm, respectively. We have chosen CoO as a model
AFM for the present study because it allows us to introduce
conveniently nonmagnetic defects at the Co sites by just controlling
the partial pressure of oxygen ($p$(O$_2$)) during the growth of the
CoO layer. The over-oxidation of CoO under high $p$(O$_2$) yields a
Co$^{2+}$-deficient layer, Co$_{1-y}$O. Thus the (intentionally)
diluted sample ($y \neq$ 0) contains a CoO layer which was grown at
a high $p$(O$_2$) (= 5$\times 10^{-6}$ mbar). On the other hand, the
CoO layer in the (nominally) undiluted ($y\rightarrow0$) sample was
grown at low $p$(O$_2$) (= 4$\times 10^{-7}$ mbar). These pressures
were carefully chosen after a number of tests and were found to
yield representative values of $m_{\rm{TRM} }$ and $B_{\rm{EB}}$ for
the respective diluted and undiluted samples.\cite{12}

The epitaxy of our samples has been established in-situ by
reflection high energy electron diffraction (RHEED). The RHEED
patterns of an undiluted Co$_{1-y}$O ($y\rightarrow0$) layer and a
diluted Co$_{1-y}$O ($y \neq$ 0) layer grown on the MgO(100)
substrate are presented in the insets (a) and (b) of Fig. 1,
respectively. The electron beam was parallel to the $[010]$
direction of the MgO(100) substrate. For all the samples the growth
of the CoO directly on the MgO(100) substrate leads to untwinned
Co$_{1-y}$O(100) layers in this system.\cite{12} For the diluted
Co$_{1-y}$O layers ($y\neq0$) (grown at $p$(O$_2$) =
$5\times10^{-6}$ mbar), the destructive interference of the fcc
lattice is removed due to some empty lattice sites caused by the
over-oxidation (dilution). Hence, additional diffraction spots
become visible which correspond to a crystalline structure with a
lattice constant in the real space about twice as large as that of
the undiluted CoO (grown at $p$(O$_2$) = 4$\times10^{-7}$ mbar).
This structure is identified as the Co$_3$O$_4$ phase which is
formed in the diluted sample due to overoxidation of Co. For AFM/FM
bilayers the Co layer grew in an hcp lattice structure with
($11\bar{2}0$)-orientation (not shown). Magnetic characterization
was performed by superconducting quantum interference device (SQUID)
magnetometry after the samples were field cooled (FC) from 340 K
through the N\'{e}el temperature ($T_{\rm{N}}$ = 291 K) to 5 K in a
field of +7 T oriented parallel to the plane of the CoO film along
its easy $[010]$ axis. For AFM-only samples the $m_{\rm{TRM} }$ was
recorded as a function of $T$ during the heating of the sample from
5 K to 340 K in the absence of an external field. For AFM/FM
bilayers $T$ was increased in steps (from 5 K to 340 K) and a
hysteresis-loop was measured between $\pm$1~T for each step. The
coercive fields of the hysteresis cycles $B_{C1}$ for descending and
$B_{C2}$ for ascending field branches were used to determine
$B_{\rm{EB}} = (B_{C1}+B_{C2}) /2$.

\begin{figure}
\includegraphics{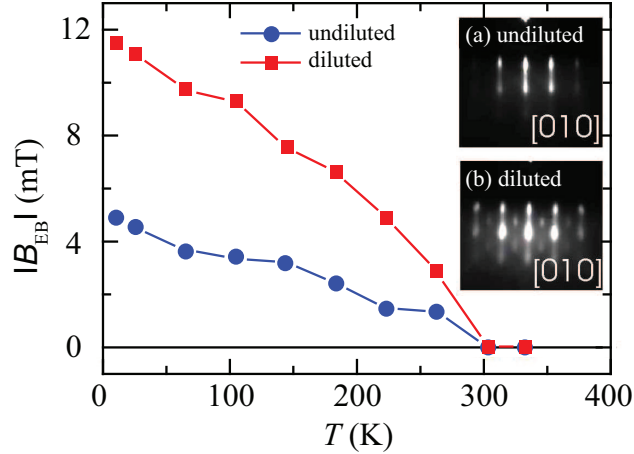}
\caption{\label{fig1} (color online) Exchange bias field
$|B_{\rm{EB}}|$ of MgO(100)/Co$_{1-y}$O(100)/Co($11\bar{2}0$)/Au vs
$T$ for undiluted (circles) and diluted (squares) samples. The inset
shows RHEED patterns of (a) the undiluted CoO layer and (b) the
diluted Co$_{1-y}$O layer grown on an MgO(100) substrate. The
electron beam direction is parallel to $[010]$ of the MgO(100)
substrate.}
\end{figure}

Figure~1 shows the $T$ dependence of $B_{\rm{EB}}$ for both
undiluted and diluted AFM/FM bilayer samples. A distinct enhancement
of $B_{\rm{EB}}$ upon dilution is evident below 291~K. However, no
change in the blocking temperature, $T_B$ (at which $B_{\rm{EB}}$ =
0), due to dilution was noticed. This is consistent with an up to
5~$\%$ dilution of Co$^{2+}$ by Mg$^{2+}$ in
Co$_{1-x}$Mg$_x$O.\cite{7} The constant $T_B$ we attribute to the
high anisotropy of CoO ($\sim 2 \times 10^7$ J/m$^3$, see, e.g.,
Ref. \cite{21}) which is an Ising-type AFM, making it more robust
against magnetic degradation upon dilution. This is in contrast to,
e.g., metallic EB systems with low \cite{22} or intermediate
anisotropy \cite{8} AFMs which show a more strongly reduced $T_B$
upon dilution.

Figure~2 shows the $T$ dependence of $m_{\rm{TRM} }$ for both FC
diluted (curve I) and FC undiluted (curve II) sole AFM samples. The
reference level is set by the zero field cooled (ZFC) diluted sample
(curve III). As expected, a strong dilution-induced enhancement
($\sim 400 \%$ at 5 K) is observed in the $m_{\rm{TRM} }$ of the FC
diluted sample in comparison to the FC undiluted one. The overall
$T$ dependence of the FC $m_{\rm{TRM} }$ of the diluted sample
compared to the undiluted one exhibits two distinct features with
decreasing $T$: (i) a monotonically increasing enhancement between
370~K and 100~K and (ii) an abrupt increase in $m_{\rm{TRM} }$ for
$T < 50$~K. The dilution-induced enhancement of the FC $m_{\rm{TRM}
}$ of sole-CoO layers above 50 K is roughly similar to the one
observed for $B_{\rm{EB}}$ of diluted CoO/Co bilayers in Fig. 1. It
is in agreement with the domain state model.\cite{7,11} However, as
opposed to $m_{\rm{TRM} }$ the entire $T$ dependence of
$B_{\rm{EB}}$ is monotonic and it lacks the abrupt increase below
50~K. The difference of $B_{\rm{EB}}(T)$ and $m_{\rm{TRM} }$(T) for
$T < 50$ K (Figs. 1 and 2, respectively) is attributed to the low
anisotropy of the uncompensated AFM spins \cite{6}, which is
insufficient to pin the FM layer. This is evidenced by the missing
strong increase of the EB field below 50~K (Fig. 1). The "isolated"
uncompensated AFM spins freeze in a $B$ field at low temperatures
($T < 50$~K), since they are weakly exchange coupled to neighboring
spins within the core of the AFM CoO due to missing or frustrated
exchange bonds. The magnetic field stabilizes the uncompensated
spins, whereas zero-field cooling does not exhibit any $m_{\rm{TRM}
}$ (see Fig. 1).

\begin{figure}
\includegraphics{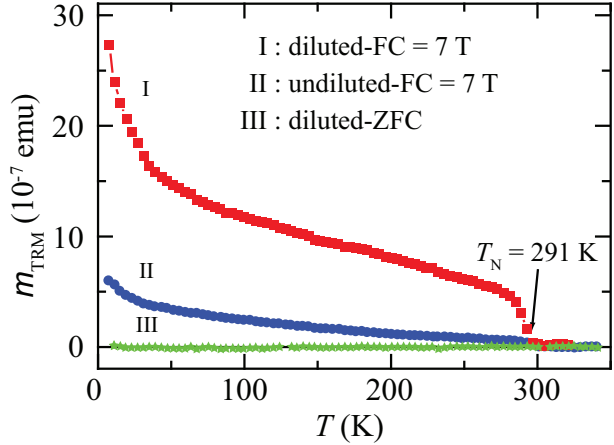}
\caption{\label{fig2} (color online) Thermoremanent magnetization of
field cooled sole-AFM MgO(100)/Co$_{1-y}$O(100)/Au as a function of
$T$ for undiluted (solid circles, II) and diluted (squares, I)
samples.  The zero field cooled (stars, III) curve of the diluted
sample is shown for reference.
}
\end{figure}

We now focus on the cycle number dependence of $m_{\rm{TRM}}^{irr}$.
A sole diluted Co$_{1-y}$O(100) sample was cooled from 340~K to 5~K
in an external field of +7~T. Subsequently, at 5~K the hysteresis
loops were measured by cycling $B$ between -7~T and +7~T. The
overall procedure is similar to the measurement of a usual
hysteresis loop of an FM. However, during each field cycle, we stop
the measurement at $B = 0$ for some time in both the decreasing and
increasing field branches. The remanent value of $m_{\rm{TRM} }$ was
then measured (Fig. 3) as a function of time ($t$) for both
ascending (lower curves) and descending (upper curves) field
branches. It is evident from Fig. 3 that the $m_{\rm{TRM} }$ is not
constant but that it decreases both as a function of time and cycle
number $n$ especially for the descending field branches.

\begin{figure}
\includegraphics{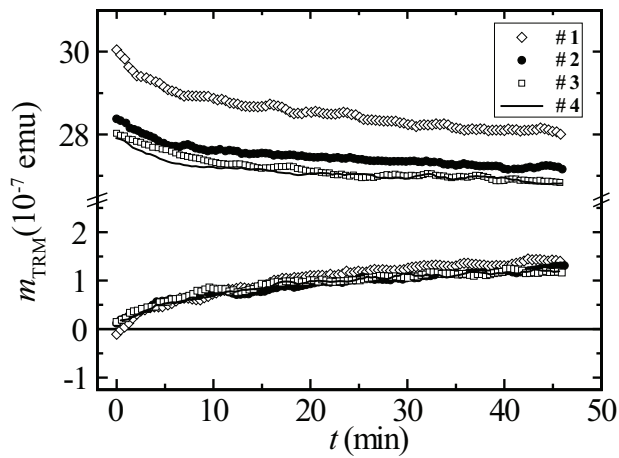}
\caption{\label{fig3} Thermoremanent magnetization at 5 K of the
diluted sole-AFM MgO(100)/Co$_{1-y}$O(100)/Au sample vs time at
$B=0$ (in remanence) for both ascending (lower curves) and
descending (upper curves) field branches of successive hysteresis
loop cycles (indicated by $\#$1 - 4). Details about the measurement
procedure are described in the text.
}
\end{figure}

For a given cycle number $n$ the mean of the values of $m_{\rm{TRM}
}$ of the upper and lower curves in Fig.~3 characterizes the
vertical shift of the \textit{hysteresis loop} of the AFM
layer.\cite{11,17} The vertical shift can be attributed to an
additional effective field on the FM, thus yielding EB. We have
calculated this mean for $t = 0$, i.e. for the time when the field
was set to zero during the AFM hysteresis loop measurement. This
quantity measures the irreversible domain state magnetization
$m_{\rm{TRM}}^{irr}$  in the whole AFM layer \cite{8,22} and is
plotted as a function of cycle number $n$ in Fig. 4 (open circles).
Clearly, the $m_{\rm{TRM}}^{irr}$ is not constant during successive
field cycles; instead it decreases monotonically during each cycle.

\begin{figure}
\includegraphics{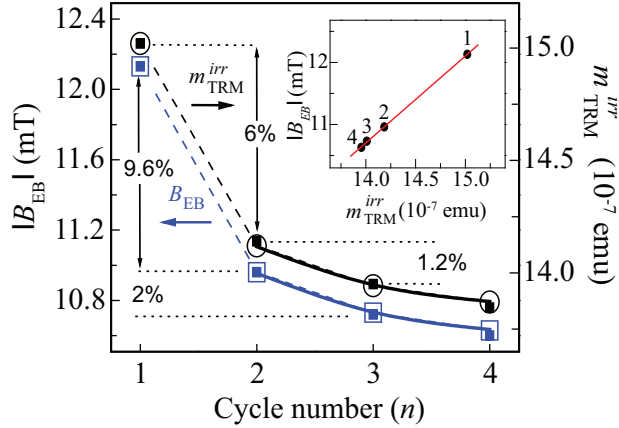}
\caption{\label{fig4} (color online) Training effect of exchange
bias field, $|B_{\rm{EB}}|$, of a diluted CoO/Co bilayer (open
squares) and of the cycle dependence of $m_{\rm{TRM}}^{irr}$ of a
diluted sole CoO layer (open circles) at 5~K. The solid lines show
fits by Eq. (1) to the data for $n > 1$. Solid squares are the
respective calculated data points generated from Eq.~2. The inset
shows $|B_{\rm{EB}}|$ vs $m_{\rm{TRM}}^{irr}$ at 5~K; the solid line
is a fit to the data points marked by their respective cycle number
$n$.
}
\end{figure}

In order to identify the origin of the EB effect we have also
plotted the TE of $B_{\rm{EB}}$ at 5 K (open squares) in Fig. 4.
This was recorded for a diluted Co$_{1-y}$O(100)/Co bilayer after
field cooling at +7~T from 340~K to 5~K. As a reference for our
SQUID measurement, we have tested undiluted CoO in a field cooled
CoO/Co bilayer at 5 K by exposing it to a reversed field of -0.5~T
during waiting times of 0 min. and 60 min. No time dependence of the
hysteresis loop was observed. In Fig. 4 a good qualitative agreement
is clearly visible between the cycle dependences of $B_{\rm{EB}}$
and of $m_{\rm{TRM}}^{irr}$ at 5~K. The maximum decrease in both
quantities occurs between the first and second field cycle and they
asymptotically approach constant values for the remaining cycles.
The following empirical formula has been widely used to describe the
TE, \cite{23}
\begin{equation}
B_{\rm{EB}} (n) - B_{\rm{EB}}(\infty) = \frac{k}{\sqrt{n}},
\end{equation}
where $k$ is a material dependent constant and $B_{\rm{EB}}(\infty)$
is the EB field in the limit of an infinite number of loops. The
solid lines in Fig. 4 show the best fits to $B_{\rm{EB}}$ and
$m_{\rm{TRM}}^{irr}$ data using Eq.~1 for $n >1$. The resulting
parameters obtained from the fit for $B_{\rm{EB}}$(n) are
$B_{\rm{EB}}(\infty) = 10.2$~mT and $k = 0.8$~mT. Similarly for
$m_{\rm{TRM}}^{irr}(n)$ the fitting parameters
$m_{\rm{TRM}}^{irr}(\infty)$ and $k'$ were found to be
$13.5\times10^{-7}$~emu and $0.6\times10^{-7}$~emu, respectively.
The fits clearly show a good agreement with the data for $n > 1$. It
should be noted that the experimental data points at $n = 1$
significantly exceed the values obtained by simple extrapolation of
the fits to $n = 1$ (not shown). The strong TE of $B_{\rm{EB}}$
between the first and second hysteresis loop has been attributed to
some initial nonequilibrium arrangement or metastable state of the
AFM spins.\cite{17,24,25,26,27,28} The exact mechanism for the
initial AFM spin arrangement is still subject to debate. Hoffmann
\cite{27} has pointed out that due to biaxial anisotropy axes in the
AFM a noncollinear arrangement of the AFM (sublattice) spins can
initially be stabilized after field cooling. This leads for
perpendicular spin arrangements to a sharp drop in the descending
field branch of the first hysteresis loop as the AFM spins relax
into a collinear arrangement. Beckmann \textit{et al.} \cite{25}
have shown that a misalignment between the cooling field direction
and the easy axis of the AFM can result in a nonequilibrium
arrangement of the AFM spins with a net $m_{\rm{TRM}}^{irr}$
oriented in a direction determined by the relative orientations
between the cooling field and AFM easy axis. During the field
cycling $m_{\rm{TRM}}^{irr}$ tends to find an energetically most
favourable orientation via irreversible rearrangements in the AFM
spin configuration. This leads to a partial loss of
$m_{\rm{TRM}}^{irr}$ and thus of $B_{\rm{EB}}$ during each cycle,
with the maximum decrease taking place during the first cycle.

Although the above dependence of the TE (Eq.~1) has been widely
observed, it lacks a physical basis. Alternatively, Binek \cite{13}
has considered the TE of AFM/FM bilayers in the thermodynamic
framework of spin configurational relaxation at the AFM surface.
This spin relaxation is activated by the consecutive cycling of the
external field. The following recursive formula is obtained for
describing the TE of $B_{\rm{EB}}$ and $m_{\rm{TRM}}^{irr}$,

\begin{equation}
  F(n+1)-F(n)=-\gamma[F(n)-F(\infty)]^3
\end{equation}

with $F$ describing $B_{\rm{EB}}$ (using $\gamma$) or
$m_{\rm{TRM}}^{irr}$ (using $\gamma'$). Taking the respective
initial values (for $n$ = 1) of $B_{\rm{EB}}$ and
$m_{\rm{TRM}}^{irr}$ as obtained from the experiment (Fig. 4), the
calculated data (solid squares in Fig. 4) are obtained from the
recursive formula in Eq.~2. For $B_{\rm{EB}}$,
 $\gamma$ and
$B_{\rm{EB}}(\infty)$ are 0.05 (mT)$^{-2}$ and 9.26 mT,
respectively. Similarly, for $m_{\rm{TRM}}^{irr}$ the parameters
$\gamma'$  and $m_{\rm{TRM}}^{irr}(\infty)$ are 0.08 (10$^{-7}$
emu)$^{-2}$ and 12.8 (10$^{-7}$ emu), respectively. Clearly, Eq.~2
(2) describes the TE of $B_{\rm{EB}}$ and of $m_{\rm{TRM}}^{irr}$
fairly well, not only for $n > 1$ but also for $n = 1$.

The inset of Fig. 4 shows a direct correlation between
$m_{\rm{TRM}}^{irr}$ and $B_{\rm{EB}}$  for the respective field
cycles marked by their number. The solid line represents the best
linear fit. The observed correlation between the TE of
$m_{\rm{TRM}}^{irr}$ and that of $B_{\rm{EB}}$ suggests that TE of
$B_{\rm{EB}}$ is due to the loss of $m_{\rm{TRM}}^{irr}$, i.e. due
to irreversible changes in the AFM domain state magnetization during
the field cycles. It should be noted that during each cycle the
percent reductions (see labelings in Fig. 4) in the respective
values of $m_{\rm{TRM}}^{irr}$ and $B_{\rm{EB}}$ do not agree
quantitatively. For example during the first field cycle
$B_{\rm{EB}}$ shows 9.6 $ \%$ reduction in comparison to the 6.0
$\%$ reduction of $m_{\rm{TRM}}^{irr}$. These differences are due to
some experimental limitations. First, for the case of AFM/FM
bilayers the interfacial AFM spins experience in addition to the
external field a strong molecular field exerted by the magnetized
FM. This results in different strengths of the effective cycling
fields on the sole AFM and on the AFM/FM bilayer. Since the
molecular fields are typically much stronger ($\sim 100$~T)
\cite{29} than externally applied fields the AFM spins in the AFM/FM
bilayer will experience a stronger effective cycling field. This
gives rise to a relatively larger percentage of decrease in
$B_{\rm{EB}}$ of CoO/Co bilayers in comparison to that of
$m_{\rm{TRM}}^{irr}$ of the sole CoO sample. Second, our measured
$m_{\rm{TRM}}^{irr}$ includes both volume as well as surface parts
of the pinned uncompensated AFM moments, whereas $B_{\rm{EB}}$ is
primarily determined by the pinned AFM moments near the FM/AFM
interface. Another factor is the uncertainty in determining the zero
of the time scale ($t = 0$) with high accuracy, i.e. when the
magnetic field is just switched off and $m_{\rm{TRM} }$ starts to
decay. Significant time (1 - 2 min) was required to reduce the field
to zero before the decay of $m_{\rm{TRM} }$ could be recorded.

In conclusion, our investigation has shown that irreversible
thermoremanent magnetization of the sole diluted Co$_{1-y}$O(100)
AFM layer exhibits systematic reductions during successive magnetic
field cycling which is consistent with the TE of the exchange bias
measured for the corresponding Co$_{1-y}$O(100)/Co($11\bar{2}0$)
bilayer. Detailed analysis shows that the TE of the exchange bias
field of the AFM/FM bilayer has its origin in the TE of
$m_{\rm{TRM}}^{irr}$ of the sole AFM layer.

S.R.A. is grateful for funding by the Higher Education Commission
(HEC), Government of Pakistan.


\begin{thebibliography}{29}


\expandafter\ifx\csname natexlab\endcsname\relax\def\natexlab#1{#1}\fi
\expandafter\ifx\csname bibnamefont\endcsname\relax
  \def\bibnamefont#1{#1}\fi
\expandafter\ifx\csname bibfnamefont\endcsname\relax
  \def\bibfnamefont#1{#1}\fi
\expandafter\ifx\csname citenamefont\endcsname\relax
  \def\citenamefont#1{#1}\fi
\expandafter\ifx\csname url\endcsname\relax
  \def\url#1{\texttt{#1}}\fi
\expandafter\ifx\csname urlprefix\endcsname\relax\def\urlprefix{URL }\fi
\providecommand{\bibinfo}[2]{#2}
\providecommand{\eprint}[2][]{\url{#2}}

\bibitem{1}
 J.  Nogu\'{e}s and I. K. Schuller, J. Magn. Magn. Mater. {\bf 192}, 203 (1999).

\bibitem{2}
A. E. Berkowitz and K. Takano, J. Magn. Magn. Mater. {\bf 200}, 552
(1999).

\bibitem{3}
 J.  Nogu\'{e}s  {\sl et al.}, Phys. Rep. {\bf 422}, 65 (2005).

\bibitem{4}
B. Dieny {\sl et al.},  Phys. Rev. B {\bf 43}, 1297 (1991).

\bibitem{5}
J. C. S. Kools, IEEE Trans. Magn. {\bf 32}, 3165 (1996).

\bibitem{6}
K. Takano {\sl et al.},   Phys. Rev. Lett. {\bf 79}, 1130 (1997).

\bibitem{peter}
P. Milt\'{e}nyi {\sl et al.},   Phys. Rev. Lett. {\bf 84}, 4224
(2000).

\bibitem{7}
J. Keller {\sl et al.},  Phys. Rev. B {\bf 66}, 014431 (2002).

\bibitem{8}
M. Fecioru-Morariu {\sl et al.},   Phys. Rev. Lett. {\bf 99}, 097206
(2007).

\bibitem{9}
L. C. Sampaio {\sl et al.},   Europhys. Lett. {\bf 63}, 819 (2003).

\bibitem{10}
R. Morales {\sl et al.},  Phys. Rev. Lett. {\bf 102}, 097201 (2009).

\bibitem{11}
U. Nowak {\sl et al.},  Phys. Rev. B {\bf 66}, 014430 (2002).

\bibitem{12}
M. R. Ghadimi, B. Beschoten, and G. Güntherodt, Appl. Phys. Lett.
{\bf 87}, 261903
        (2005).

\bibitem{13}
 C. Binek, Phys. Rev. B {\bf 70}, 014421 (2004).

\bibitem{14}
S. Brems, K. Temst, and C. Van Haesendonck, Phys. Rev. Lett. {\bf
99}, 067201 (2007).


\bibitem{15}
P. Y. Yang {\sl et al.},  Appl. Phys. Lett. {\bf 92}, 243113 (2008).

\bibitem{16}
A. G. Biternas, U. Nowak, and R. W. Chantrell, Phys. Rev. B {\bf
80}, 134419 (2009).

\bibitem{17}
A. G. Biternas, R. W. Chantrell, and U. Nowak, Phys. Rev. B {\bf
82}, 134426 (2010).

\bibitem{18}
P. Kappenberger {\sl et al.},  Phys. Rev. Lett. {\bf 91}, 267202
(2003).

\bibitem{19}
H. Ohldag {\sl et al.},  Phys. Rev. Lett. {\bf 91}, 017203 (2003).

\bibitem{20}
A. Hochstrat, C. Binek, and W. Kleemann, Phys. Rev. B {\bf 66},
092409 (2002).

\bibitem{21}
J. Kanamori, Prog. Theor. Phys. {\bf 17}, 197 (1957).

\bibitem{22}
C. Papusoi {\sl et al.},  J. Appl. Phys. {\bf 99}, 123902 (2006).

\bibitem{23}
D. Paccard {\sl et al.},  Phys. Status Solidi {\bf16}, 301 (1966).

\bibitem{24}
D. Suess {\sl et al.},  Phys. Rev. B {\bf 67}, 054419 (2003).

\bibitem{25}
B. Beckmann, U. Nowak, and K. D. Usadel, Phys. Rev. Lett. {\bf 91},
187201 (2003).

\bibitem{26}
F. Radu {\sl et al.},  Phys. Rev. B {\bf 67}, 134409 (2003).

\bibitem{27}
A. Hoffmann, Phys. Rev. Lett. {\bf 93}, 097203 (2004).

\bibitem{28}
T. Hauet {\sl et al.},  Phys. Rev. Lett. {\bf 96}, 067207 (2006).

\bibitem{29}
C. Kittel, Introduction to Solid State Physics (Wiley, New York,
1996).


\end{thebibliography}
\end{document}